\documentclass[journal]{IEEEtran}
\IEEEoverridecommandlockouts
\usepackage{epsfig}
\usepackage[T1]{fontenc}
\usepackage[latin1]{inputenc}
\usepackage{graphicx}
\usepackage{amssymb}
\usepackage{amsmath}
\usepackage{dcolumn}
\usepackage{bm}
\usepackage{color}
\usepackage{epstopdf}
\usepackage{algorithm}
\usepackage{algorithmicx}
\usepackage{algpseudocode}
\usepackage{subfigure}
\usepackage{amssymb}
\usepackage{amsmath}
\usepackage{dcolumn}
\usepackage{bm}
\usepackage{color}
\usepackage{colortbl}
\usepackage{multicol}
\usepackage{algorithm}
\usepackage{algorithmicx}
\usepackage{algpseudocode}
\usepackage{multirow}
\usepackage{chngpage}
\usepackage{array}
\usepackage{subfigure}
\usepackage{threeparttable}
\usepackage{mathrsfs}
\usepackage{makecell}

\usepackage{mathptmx} 
\usepackage{amssymb}  
\usepackage{cite}
\usepackage[numbers,sort&compress]{natbib}

\usepackage{url}
\newcommand{\EQ}{\begin{eqnarray}}

\newcommand{\EN}{\end{eqnarray}}
\newcommand{\EQQ}{\begin{eqnarray*}}
\newcommand{\ENN}{\end{eqnarray*}}
\newcommand{\overlineray}{\begin{array} }

\newcommand{\R}{{\mathbb R}}
\newcommand{\C}{{\mathbb C}}
\newcommand{\barray}{\begin{array} }
\newcommand{\earray}{\end{array}}
\renewcommand{\t}{^{\mbox{\tiny\sf T}}}
\newcommand{\bremark}{\begin{remark} }
\newcommand{\eremark}{\end{remark}}
\newcommand{\btheorem}{\begin{theorem}}
\newcommand{\etheorem}{\end{theorem}}
\newcommand{\blemma}{\begin{lemma}}
\newcommand{\elemma}{\end{lemma}}
\newcommand{\bassumption}{\begin{assumption} }
\newcommand{\eassumption}{\end{assumption}}
\newcommand{\bcorollary}{\begin{corollary} }
\newcommand{\ecorollary}{\end{corollary}}
\newcommand{\bdefinition}{\begin{definition} }
\newcommand{\edefinition}{\end{definition}}
\newcommand{\bproposition}{\begin{proposition}}
\newcommand{\eproposition}{\end{proposition}}

\newcommand{\balgorithm}{\medskip\begin{algorithm} \rm}
\newcommand{\ealgorithm}{ \hfill \rule{1mm}{2mm}\medskip
\end{algorithm} }
\newtheorem{remark}{\rm\bfseries Remark}[section]
\newtheorem{corollary}{\rm\bfseries Corollary}[section]
\newtheorem{definition}{\rm\bfseries Definition}[section]
\newtheorem{theorem}{\rm\bfseries Theorem}[section]
\newtheorem{lemma}{\rm\bfseries Lemma}[section]
\newtheorem{assumption}{\rm\bfseries Assumption}[section]
\newtheorem{proposition}{\rm\bfseries Proposition}[section]

\usepackage{nomencl}
\makenomenclature
\usepackage{etoolbox}
\renewcommand\nomgroup[1]{%
  \item[\bfseries
  \ifstrequal{#1}{A}{Notations}{%
  \ifstrequal{#1}{B}{Parameters}{}}%
]}

\begin{document}

\title{Probabilistic Optimal Power Flow Considering Correlation of Wind Farms via Markov Chain Quasi-Monte Carlo Sampling}

\author{Weigao Sun, \textit{Student Member}, \textit{IEEE}, Mohsen Zamani,  Hai-Tao Zhang, \textit{Senior Member}, \textit{IEEE}, Yuanzheng Li
\thanks{This work was supported by the National Natural Science Foundation of China under Grant U1713203 and 51328501. (Corresponding authors: Mohsen Zamani and Hai-Tao Zhang.)

W. Sun, H.-T. Zhang and Y. Li are with the School of Automation, Huazhong University of Science and Technology, Wuhan 430074, P.R.~China (email: \{sunweigao, zht, yuanzheng\_li\}@hust.edu.cn).

Mohsen Zamani is with the School of Electrical Engineering and Computing, University of Newcastle, University Drive, Callaghan, NSW 2308, Australia (email: mohsen.zamani@newcastle.edu.au).
}}

\maketitle

\IEEEpeerreviewmaketitle

\begin{abstract}
The probabilistic characteristics of daily wind speed are not well captured by simple density functions such as Normal or Weibull distribuions  as suggested by the existing literature. The unmodeled uncertainties can cause unknown influences on the power system operation. In this paper, we develop a new stochastic scheme for the probabilistic optimal power flow (POPF) problem, which can cope with arbitrarily complex wind speed distributions and also take into account the correlation of different wind farms. A multivariate Gaussian mixture model (GMM) is employed to approximate actual wind speed distributions from multiple wind farms. Furthermore, we propose to adopt the Markov Chain Monte Carlo (MCMC) sampling technique to deliver wind speed samples as the input of POPF. We also novelly integrate a Sobol-based quasi-Monte Carlo (QMC) technique into the MCMC sampling process to obtain a faster convergence rate. The IEEE 14- and 118-bus benchmark systems with additional wind farms are used to examine the effectiveness of the proposed POPF scheme.

\end{abstract}



\nomenclature[A]{$\R$}{The set of real numbers.}
\nomenclature[A]{$\R^+$}{The set of positive real numbers.}
\nomenclature[A]{$\C$}{The set of complex numbers.}
\nomenclature[A]{$\top$}{The matrix transposition operator.}
\nomenclature[A]{$\mathcal{N}(\cdot)$}{Normal distribution function.}
\nomenclature[A]{$N_b := \{1, \cdots, n_b\}$}{The set of all buses.}
\nomenclature[A]{$N_g := \{1, \cdots, n_g\}$}{The set of traditional generator buses.}
\nomenclature[A]{$N_w := \{1, \cdots, n_w\}$}{The set of buses with wind farms.}
\nomenclature[A]{$N_l := \{1, \cdots, n_l\}$}{The set of load buses.}
\nomenclature[A]{$E := \{1, \cdots, e\}$}{The set of transmission lines.}
\nomenclature[B]{$V_{i}, \theta_i$}{Voltage magnitude $V_{i}$ and voltage angle $\theta_i$ on bus $i$.}
\nomenclature[B]{$Y_{ij} := G_{ij}+ jB_{ij}$}{The admittance from bus $i$ to bus $j \in N_b$, includes conductance $G_{ij} \in \R^{n_b\times n_b}$ and susceptance $B_{ij} \in \R^{n_b\times n_b}$.}
\nomenclature[B]{$P^G_k$, $Q^G_k$}{The active and reactive power on traditional generator $k \in N_g$.}
\nomenclature[B]{$P^W_j$, $Q^W_j$}{The active and reactive power on wind farm $j \in N_w$.}
\nomenclature[B]{$P^D_i$, $Q^D_i$}{The active and reactive load power on bus $i \in N_l$.}
\nomenclature[B]{$P_{cd}$, $S_{cd}$}{The active and apparent power from bus $c \in N_b$ to the rest of grid through branch $(c,d)\in N_l$. Here $d \in N_b$.}
\nomenclature[B]{$V_{i}^{\min}$, $V_{i}^{\max}$}{Constant constraint boundaries of voltage magnitude on bus $i \in N_b$.}
\nomenclature[B]{$P_k^{G, \min}$, $P_k^{G, \max}$}{Constant constraint boundaries of active power on generator $k \in N_g$.}
\nomenclature[B]{$Q_k^{G, \min}$, $Q_k^{G, \max}$}{Constant constraint boundaries of reactive power on generator $k \in N_g$.}
\nomenclature[B]{$P_j^{W, \min}$, $P_j^{W, \max}$}{Constant constraint boundaries of active power on wind farm $j \in N_w$.}
\nomenclature[B]{$Q_j^{W, \min}$, $Q_j^{W, \max}$}{Constant constraint boundaries of reactive power on wind farm $j \in N_w$.}
\nomenclature[B]{$P_{cd}^{\max}$, $S_{cd}^{\max}$}{Constant constraint boundaries of active and apparent power on line $(c,d)\in N_l$.}
\nomenclature[B]{$\triangle V_{cd}^{\max}$}{Constant constraint boundaries of voltage magnitude on line $(c,d)\in N_l$.}
\printnomenclature[3cm]

\section{Introduction}
\label{sec: intro}

Wind power generation {is experiencing} tremendous developments as a clean and renewable energy {resource} \cite{Wang2016}. It considerably contributes to the long-term sustainability of power systems, {whereas}, introduces {significant} uncertainties {into the overall networks} as well. {Uncertainty involved with wind power generation} may cause {operational} problems, such as {overload of transmission lines}, which {in turn} threaten the {reliability and security} of power system \cite{Chen2015, Wang2018}. {It is challenging to study how the uncertainty
involved with wind power generation will influence power
system operations.} {Therefore}, {probabilistic optimal power flow (POPF)}, as a powerful tool to analysis uncertainties, has attracted considerable attention \cite{Li2014, Ke2016, Zhang2017, Xie2018, Kazemdehdashti2018, Li2008, Schellenberg2005, Zou2014, Aien2014, Verbic2006, Schellenberg2005_1}. {Instead} of calculating the {traditional} deterministic optimal power flow (DOPF) \cite{Wei2018}, POPF treats each uncertain variable in power systems as a random variable with certain probabilistic distribution and aims to obtain the statistical information of the optimal solutions. By evaluating the statistical information of output variables, e.g., mean, stardand deviation or even probabilistic density function, it is {promising} to {figure out} the potenial risk and weakness of the power system under investigation.

%
%

{The existing literature concerning} the POPF problem falls into three categories: {analytical-based methods} \cite{Schellenberg2005, Li2008, Schellenberg2005_1}, {the} point estimation \cite{Aien2014, Li2014, Verbic2006, Su2005} and Monte Carlo {(MC)} {simulations} \cite{Xie2018, Yu2009, Hajian2013, Xu2017}. Considering the {large} computation {burden} of solving POPF, analytical methods were developed first. The {essential} idea of analytical {methods} is {to compute} statistical moments of output variables in POPF based on the moments {associated with} input variables. {For instance}, \cite{Schellenberg2005} developed a cumulant method for POPF problem, {which} {assumed the relationship between linear input and cumulant output}. {Then, if the} input variables follow Gaussian or Gamma distributions, which {are} analytical known, cumulant method {could be} {deployed} to obtain statistical {solutions} of output variables. In \cite{Li2008}, the authors proposed a {method that exploits} the first-order Taylor series expansion, {therein} the first two moments of the input load power {could be used to} obtain the statistical information of the output variables. Analytical methods like \cite{Schellenberg2005} and \cite{Li2008} compute swiftly, however, suffer from the issue of accuracy. The {implementation} of analytical methods in practice also {depends heavily} on the particular optimal power flow formulations.

Point estimation method has been {adopted} to solve the POPF problem as well as the probabilistic power flow (PPF) problem \cite{Yu2009, Hajian2013, Wang2017a, Fan2012, Williams2013, Xu2017}. In \cite{Aien2014} and \cite{Verbic2006}, a two-point estimation method was proposed for POPF problem. It is worth mentioning that in \cite{Aien2014}, the correlation of input variables {was described by the} coefficients matrix. The reference \cite{Li2014} {addressed} the POPF problem {by} {calculating} correlation of wind speeds {via} improving the point estimation method. {However}, the point estimation {scheme} only calculates first few statistical moments, {which is not accurate enough}. {Meanwhile, the} computational burden is proportional to the {uncertain variables numbers}, which {hinders the further applications} in large-scale power systems.

{In addition, MC methods have} been widely studied for PPF and POPF problems. With the samples from probabilistic density of input variables, the deterministic power flow or optimal power flow is calculated repeatly which generates samples of output variables. Routine {MC} method with large enough repeating times (e.g. 10000 times) can give {sufficiently} accurate results. However, {it is} {computationally} {expensive}. To solve such a dilemma, improved sampling methods were employed to reduce the computational burden. {For instance, Latin hypercube sampling (LHS) \cite{Yu2009}, Latin supercube sampling \cite{Hajian2013} and quasi-MC (QMC) methods \cite{Xie2018}\cite{Xu2017} are representative works}.

Correlation of the random input variables in PPF or POPF problem has {attracted} {much} attention in recent years. {Ref.} {\cite{Aien2014} adopted} the correlation matrix technique into the point estimation method to {address} the correlation of wind power generation and loads. {Ref.} {\cite{Li2014} studied} correlation of wind speeds with different distributions. Copula function was {utilized} in \cite{Xie2018} to {describe} the dependent structure of random wind speeds. In \cite{Singh2010}, a Gaussian mixture model (GMM) was proposed to {approximate} the probabilistic distribution of loads. In \cite{Ke2016} and \cite{Wang2017}, the multivariate GMM was {adopted} to describe the wind power uncertainties and their correlation.

\begin{figure}[htbp]
\centering
{\includegraphics[width=9cm]{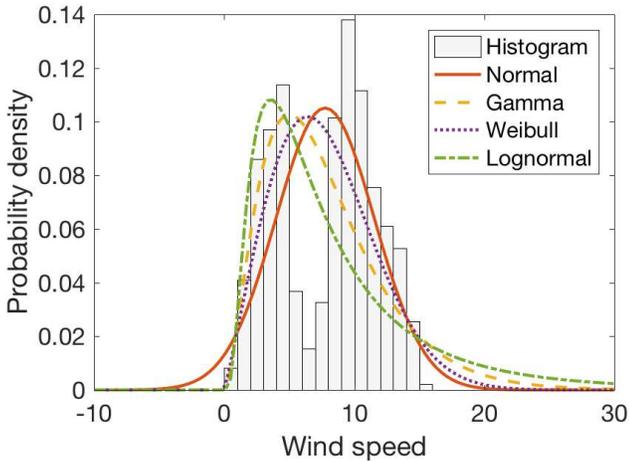}}
\caption{Distribution of actual daily wind speed.}
\label{fig: fig_uncertainty_wind}
\end{figure}

{As we will show later in the paper, the real world daily wind speed data obeys unknown complex distributions, not Weibull distribution which is commonly used to model the probabilistic property of wind speed, see e.g., \cite{Li2014} and \cite{Ke2016}. In this paper, we develop a novel stochastic scheme for solving POPF which has no such presumption on the probability density function associated with the collected data. Compared to the existing relevant state-of-the-art works, the proposed scheme can cope with arbitrarily complex wind speed distributions and also take into account the correlation of different wind farms with the help of  multivariate GMM. This is the first contribution of this work. Given complications arise from sampling multivariate GMM, the second contribution lies in adopting the powerful Markov chain Monte Carlo (MCMC) sampling technique to obtain inout samples for POPF and also   novelly integrating Sobol-based QMC method into the MCMC sampling process for a faster convergence rate.}

The paper is organized as follows. Section~\ref{sec: preliminary} {provides a preliminary {of} the probabilistic characteristics of power system uncertainties, and {thereby} presents} the POPF problem. Section~\ref{sec: gmm} introduces the multivariate GMM to represent the joint distribution of wind speeds in multiple wind farms. In Section~\ref{sec: sampling}, An MCMC sampler is given and then improved by integrating the sobol-based QMC technique. Section~\ref{sec: prodedure} summarizes the proposed scheme for solving POPF problem. The present scheme is {examined} by case studies in Section \ref{sec: case study}. Finally, conclusion is drawn in Section~\ref{sec: Conclusions}.

\section{{Preliminary}}
\label{sec: preliminary}

{We review the sources of uncertainties in power systems and then introduce the POPF problem in this section.}

\subsection{{Uncertainties in Power Systems}}

The power system {considered} in this paper can be sketched as a network represented by a connected undirected graph ($N$, $E$), {where} $N$ and $E$ are {sets} of buses and branches in a power system, respectively. Bus 0 is the slack bus, {without loss of generality}, its voltage phase angle is taken as reference and set as zero \cite{Tang2017}. Other buses are classified into generator {or} load buses. Here, we consider a power system {including both} {traditional thermal and wind power generations}.

The uncertainties of such a power system mostly {stems} from the randomness {associated with} load and wind speeds. The load, as an uncertain variable, is often influenced by the {usage} time, the market electrical price and even the weather condition. It is a common practice to model the probabilistic distribution of load as a normal distribution with parameters {obtained} from historical data. {Thereby}, we {follow the same remedy and} {describe} the load as a normal distribution {and set} its mean equal to the base load, and its standard deviation to $5\%$ of its mean \cite{Aien2014}.

The randomness of wind power generation also plays an important role in the uncertainties of power systems. Wind speeds {vary} with the time, weather and the location of wind farm {and this in turn} {results in} {variation} of wind power generation. The probabilistic distribution of wind speed is commonly claimed to follow the Weibull distribution in literatures \cite{Patel2005}. However, it {could also obey} the Burr or lognormal distribution, or the {combination} of Weibull, Burr and lognormal distributions, as {claimed} in \cite{Li2014}. {Indeed, the distribution of real daily wind speed can be arbitrarily complex instead of following several simple distributions, as shown in Fig. \ref{fig: fig_uncertainty_wind}. Here, the wind speed data is collected from the Measurement and Instrumentation Data Center under the National Renewable Energy Laboratory\footnote{\url{http://midcdmz.nrel.gov/}}.}

{A mapping for transforming wind speed to electrical power in} a wind turbine is typically described as:
\begin{equation}
  \label{eq: speed 2 power}
P_{t}(v)=
\begin{cases}
0,& v\leq v_{\text{in}},\\
f(v),& v_{\text{in}}\leq v \leq v_{\text{r}},\\
P_r,& v_{\text{r}}\leq v \leq v_{\text{out}},\\
0,& v\geq v_{\text{out}},
\end{cases}
\end{equation} where {$v_{\text{in}}$, $v_{\text{out}}$, $v_{\text{r}}$} are the cut-in, cut-out and rated wind speed, respectively. And $P_{t}$ is the real power generation of a wind turbine. Here $f(v)$ represents the generation mechanism of the wind turbine in standard working ranges \cite{Xie2018}.
The real power $P^W$ and reactive power $Q^W$ generated by wind trubines in the $j$-th wind farm are {shown as follows}
\EQQ
P^W_j &=& P_{t}\cdot N_{t},\\
Q^W_j &=& \frac{P^W}{\cos \varphi} \cdot \sqrt{1-\cos^2 \varphi},
\ENN where $N_{t}$ is the number of wind turbines in {a wind farm} and $\cos \varphi$ is the power factor.


\subsection{{POPF}}
{We {now} present the formulation of {POPF} problem. In a power system, the variables in DOPF {fall} into two categories: {the control variables $\bm{u}=[P^G, V_i|i\in N_g]$ and the state variables $\bm{x}=[Q^G, \theta, V_j| j\in N_l]$.} The DOPF aims to find the minimum of power generation cost by adjusting the control variables $\bm{u}$ subject to the power flow equations and other security constraints.} The DOPF problem
solves the nonlinear constrained optimization problem:
\EQ \label{eq: opf formulation f}
\min_{\bm{u}} f (\bm{u})
\EN subject to
\EQQ
P^G_i+P^{W}_i-P^D_i = \sum_{j=1}^{n_b} V_{i}V_{j}(G_{ij}\cos\theta_{ij}+B_{ij}\sin\theta_{ij}), \\
Q^G_i+Q^{W}_i-Q^D_i = \sum_{j=1}^{n_b} V_{i}V_{j}(G_{ij}\sin\theta_{ij}-B_{ij}\cos\theta_{ij}),
\ENN
\EQQ
V_{i}^{\min}\leq V_{i}\leq V_{i}^{\max}, \\
P_k^{G, \min}\leq P_k^G\leq P_k^{G, \max}, \\
Q_k^{G, \min}\leq Q_k^G\leq Q_k^{G, \max}, \\
P_j^{W, \min}\leq P_j^W\leq P_j^{W, \max}, \\
Q_j^{W, \min}\leq Q_j^W\leq Q_j^{W, \max}, \\
|S_{cd}|\leq S_{cd}^{\max}, \\
|P_{cd}|\leq P_{cd}^{\max}, \\
V_{c}-V_{d}\leq \triangle V_{cd}^{\max},
\ENN
where {$\forall i \in N_b$, $\forall k \in N_g$, $\forall j \in N_w$, $\forall (c,d)\in N_l$} and {$f(\cdot)$ is in general a convex polynomial objective function, see e.g. \cite{Tang2017} {and} \cite{Gan2014}. In this paper, we set $f(\cdot)$ as:
\EQQ
f(P^G)&=&\sum_{k=1}^{n_g} {f_k(P^{G}_k)}+\sum_{j=1}^{n_w}d_j\\
&=&\sum_{k=1}^{n_g} (a_k+b_k P^{G}_k+ c_k (P^{G}_k)^2)+\sum_{j=1}^{n_w}d_j,\ENN
where $a_k,~b_k~\text{and}~c_k$ are the constant, linear and quadratic coefficient of the cost of $k$-th traditional thermal generator, respectively; $d_j$ is the {constant} maintenance cost of $j$-th wind farm.}

{The minimization problem (\ref{eq: opf formulation f}) and its associated nonlinear constraints can be written into {a compact form as}
\EQ
Z=h(W),
\label{eq: POPF compact form}
\EN
with $W = [P^G, V_i, P^W, Q^W, P^D, Q^D|i\in N_g]$. {The parameter} $Z$ captures those variables of interest, e.g., the generation cost, bus voltage, active and reactive power flow. In order to capture the uncertainty involved in generation of wind power, we presume that $P^W$, $Q^W$ follow certain unknown probability distribution {functions}.}
By using sampling techniques, we obtain samples from the {underlying} distributions {associated with those} random variables in $W$. Then the DOPF is performed {recursively} with the samples of {these variables}, which {in turn yields {output samples}. One can exploit these samples information to attain estimation about statistical properties of these variables of interest.} {Afterwards, evaluating} {these} statistical information, it is possible to find the potenial risk and the weakness of the power system under investigation. For example, {such analysis} can provide the probabilities of line overloading or bus overvoltage.


\section{Multivariate Gaussian Mixture Model}
\label{sec: gmm}
{To approximate wind speed distribution, we will seek assistance from the GMM, which} is a probabilistic model that assumes all the data points are generated from a mixture of a finite number of Gaussian distributions with unknown parameters \cite{Ross2014}. Each Gaussian distribution is called a Gaussian component. {GMM can capture} arbitrarily complex distributions by using specific number of Gaussian components with different parameters \cite{Robert2014}. {Therefore}, GMM has been widely used in data classification and machine learning. {Meanwhile, it} has been {verified} to be able to model the uncertainties of power systems {in \cite{Ke2016}, \cite{Singh2010} and \cite{Wang2017}.}

A GMM is called multivariate if each {of its components} is multi-dimensional Gaussian distribution. Given a random vector $\bm{x}=[\bm{x}_1, \bm{x}_2, \cdots, \bm{x}_D]\t$, the joint probability density function (PDF) of multivariate GMM is
\EQ
p(\bm{x})&=&\sum_{m=1}^M\frac{c_m}{(2\pi)^{D/2}|\bm{\Sigma}_m|^{\frac{1}{2}}}\exp[-\frac{1}{2}(\bm{x}-\bm{\mu}_m)\t\bm{\Sigma_m}^{-1}(\bm{x}-\bm{\mu}_m)]\cr
&=&\sum_{m=1}^M c_m \mathcal{N}(\bm{x}|\bm{\mu}_m, \bm{\Sigma}_m),
\label{eq: GMM PDF}
\EN where $M$ is the number of Gaussian components, which is {set as} a priori according to probabilistic characteristics of data. Random vector $\bm{x}\sim \mathcal{N}(\bm{\mu}_m\in\R^D, \bm{\Sigma}_m\in \R^{D\times D})$, here $\mathcal{N}(\cdot)$ denotes the $D$-dimensional Gaussian distribution function, $\bm{\mu}_m$ and $\bm{\Sigma}_m$ {are} the expectation and covariance matrix of $m$-th $D$-dimentional Gaussian component, respectively. The positive mixture weights $c_m$ sum to unity, {i.e.}, $\sum_{m=1}^M c_m = 1$.


The multivariate GMM in (\ref{eq: GMM PDF}) has a {parameter set} $\Theta=\{c_m, \bm{\mu}_m, \bm{\Sigma}_m | m=1,2,\cdots,M\}$ to be determined. Estimating the parameter set $\Theta$ based on historical data is known as a learning process. Here, we focus on the expectation maximization (EM) algorithm for GMM parameter estimation. EM is an iterative procedure for maximum likelihood parameter estimation from dataset with latent variables. To estimate the parameter set $\Theta$, we write the log-likelihood function as
\EQQ
l(\Theta)= \log \{\sum_{m=1}^M c_m \mathcal{N}(\bm{x}, \bm{z}|\bm{\mu}_m, \bm{\Sigma}_m)\},
\ENN where $\bm{x}$ is the observed data and $\bm{z}$ is the unobserved latent data.
{EM algorithm contains two main steps: $E$-step, it guesses the values of latent data $\bm{z}$; $M$-step, it {assumes} the gausses of $\bm{z}$ are correct, and applies the maximum likelihood estimation to update $\Theta$.}
The EM algorithm which is widely used and proved to be effective in practice {has been embedded in many commercial softwares}.


%
%
%

\section{{Sampling Method}}
\label{sec: sampling}
{To implement the POPF calculation, efficient sampling technique should be designed to obtain {sufficient} wind power samples as inputs of problem (\ref{eq: POPF compact form}).
Due to the complications arise from sampling a multivariate GMM, we propose to exploit the MCMC sampling technique to yield sufficient wind speed samples, then transform them to wind power samples by the mapping Eq. (\ref{eq: speed 2 power}).} {Furthermore,} a sobol-based QMC technique is integrated into the MCMC sampling to obtain faster convergence rate.

\subsection{MCMC Sampler}

{MCMC is a powerful sampling technique which can provide samples from arbitrary probability density $p(\bm{x})$}, which has been adopted for wind power simulation in \cite{Papaefthymiou2008}.
It works in two stages: proposal and acceptance.
Given $x_k$, a candidate point $\xi_k$ is drawn from a proposal distribution $q(\xi_k|\bm{x}_{k-1})$, where $\xi_k \sim q(\cdot|\bm{x}_{k-1})$ is a possible realization for $x_k$. Then compute the acceptance probability
\EQ
\alpha(\xi_k|\bm{x}_{k-1})=\min\Big\{1, \frac{p(\xi_k)q(\bm{x}_{k-1}|\xi_k)}{p(\bm{x}_{k-1})q(\xi_k|\bm{x}_{k-1})}\Big\},
\label{eq: acceptance probability of Metropolis Hastings}\EN
and draw a random variable $\widehat{z}$ from uniform disribution $\mathcal{U}(0,1)$. If $\widehat{z}<\alpha(\xi_k|\bm{x}_{k-1})$, accept $\xi_k$ and set $\bm{x}_k=\xi_k$. Otherwise, reject $\xi_k$ and set $\bm{x}_k=\bm{x}_{k-1}$.

MCMC generates a Markov chain $(\bm{x_0}, \bm{x_1}, \cdots, \bm{x_t}, \cdots)$, as the transtion probabilities from $\bm{x_t}$ to $\bm{x_{t+1}}$ depends only on $\bm{x_t}$. After a sufficient burn-in period, for example, $k$ steps, the Markov chain approaches its stationary distribution. Then, the samples in $(\bm{x_{k+1}}, \cdots, \bm{x_{k+n}})$
are the samples from $p(\bm{x})$, where $n$ is the number of samples.

Choice of the proposal distribution $q(\xi_k|\bm{x}_{k-1})$ has a {significant} influence on the performance of the MCMC sampler. A widely used proposal distribution is {obtained from} the random walk below:
\EQQ
\xi_k = \bm{x}_{k-1} + \nu_k,
\ENN where $\nu_k$ is a random perturbation {which is commonly a white noise. Hence} the proposal distribution $\xi_k$ is symmetric
\EQQ
q(\xi_k|\bm{x}_{k-1}) = q(\bm{x}_{k-1}|\xi_k).
\ENN In this specific situation, the acceptance probability in (\ref{eq: acceptance probability of Metropolis Hastings}) becomes
\EQ
\alpha(\xi_k|\bm{x}_{k-1})=\min\Big\{1, \frac{p(\xi_k)}{p(\bm{x}_{k-1})}\Big\}.
\label{eq: acceptance probability of Metropolis}
\EN

\begin{algorithm}[t]
\caption{QMC-MCMC Sampler}
\textbf{Step 1.} Initialize $\bm{x}_0$ satisfying $p(\bm{x}_0)>0$, and set $k=1$.

\textbf{Step 2.} At iteration $k$, draw a candidate point $\xi_k$ by Sobol-based QMC method from a proposal distribution $q(\xi_k|\bm{x}_{k-1})$ , where $\xi_k \sim q(\cdot|\bm{x}_{k-1})$ is a possible realization for $x_k$.

\textbf{Step 3.} Compute the acceptance probability
\EQQ
\alpha(\xi_k|\bm{x}_{k-1})=\min\Big\{1, \frac{p(\xi_k)q(\bm{x}_{k-1}|\xi_k)}{p(\bm{x}_{k-1})q(\xi_k|\bm{x}_{k-1})}\Big\}.
\ENN

\textbf{Step 4.} Draw a random variable $\widehat{z}$ by Sobol-based QMC method from uniform disribution $\mathcal{U}(0,1)$. If $\widehat{z}<\alpha(\xi_k|\bm{x}_{k-1})$, accept $\xi_k$ and set $\bm{x}_k=\xi_k$. Otherwise, reject $\xi_k$ and set $\bm{x}_k=\bm{x}_{k-1}$.

\textbf{Step 5.} Set $k=k+1$ and return to step 2.
\label{algo: MCMC}
\end{algorithm}

The {Eq.} (\ref{eq: acceptance probability of Metropolis}) provides an intuitive explanation of the MCMC sampler. {By this means,} $\bm{x}_k$ converges to the target distribution $p(\bm{x})$ by accepting or rejecting a candidate proposal $\xi_k$. If the proposal $\xi_k$ is more likely to be a realization of the density $p(\bm{x}_k)$ than the previous iteration $\bm{x}_{k-1}$, i.e., $p(\xi_k)>p(\bm{x}_{k-1})$, then the proposal $\xi_k$ is accepted as a realization $\bm{x}_k=\xi_k$. {Otherwise}, it still has a chance to be retained. In fact, the proposal $\xi_k$ less likely than $\bm{x}_{k-1}$ is retained with a probability {${p(\xi_k)}/{p(\bm{x}_{k-1})}\leq 1$}.


\subsection{Improving MCMC by Integrating QMC Method}

\begin{figure}[htbp]
\centering
{\includegraphics[width=9cm]{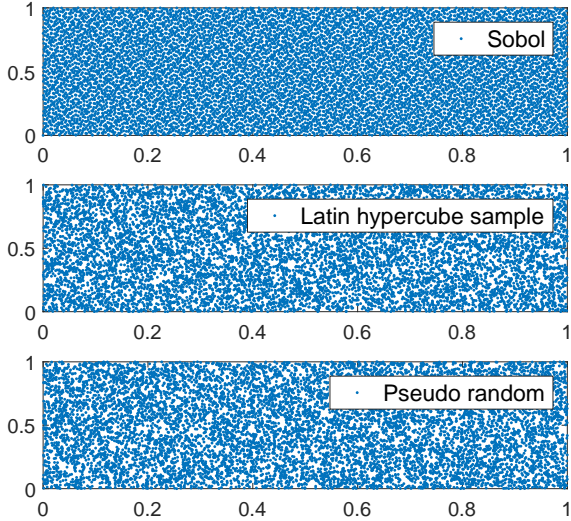}}
\caption{ 10000 uniform random numbers generated by Sobol, Latin hypercube sample and pseudo-random number technique in a 2-dimensional unit space.}
\label{fig: quasi plot}
\end{figure}

{MCMC sampler works well for sampling from arbitrary probability density functions, however, it suffers from the costly computation burden and slow convergence rate \cite{Gilks1995}. This motivates us to integrate QMC into MCMC to obtain faster convergence rate \cite{Gilks1995}. Consider a quantity $\mu$, which is of interest, can be expressed as $E(f(X))$ for a real valued function $f(\cdot)$ and random vector X with probability density $p(\cdot)$ on $\R^d$. Then, $\mu$ can be expressed as $\int_{\R^d}f(x)p(x)dx$. In simple MC, one estimates $\mu$ by
\EQQ
\widehat{\mu}_n = \frac{1}{n}\sum^n_{i=1}f(x_i),
\ENN where
$\bm{x}_i$ for $i=1,\cdots, n$ are independent random samples obtained from $p(\cdot)$. A pseudo-random number generator is commonly used to simulate the $x_i$ values, which is known as the simple random sampling (SRS).
QMC is a variant of simple MC to obtain a {higher} rate of convergence by using low-discrepancy sequences \cite{Dick2013}. In QMC, the $x_i$ values chosen deterministically are more uniformly distributed than the pseudo-random numbers in SRS.} {There {exist} several different ways to generate quasi-random low-discrepancy sequences which result in different instances of QMC method, such as the Sobol sequence \cite{Xie2018} and Latin hypercube sample \cite{Yu2009}. Here, we focus on the Sobol sequence {whose} implementation is {introduced} in \cite{Bratley1988}.} Fig. \ref{fig: quasi plot} {depicts} 10000 uniform random numbers generated by Sobol technique, Latin hypercube sample and pseudo-random number technique. {It is observed that} Sobol sequence results in more uniformly distributed (or low-discrepancy) {samples} than other {ones}. {We now discuss the convergence rate of QMC compare to MC method. To this end, we recall the follow definition.
\bdefinition {\it (Star discrepancy \cite{Niederreiter1992})}
\label{def: star discrepancy}
Let $\delta(a)=\text{Vol}([0,a])-\frac{1}{n}\sum_{i=1}^n I_{x_i\in[0,a]}$ be the local discrepancy function at point $a\in [0,1]^d$. Here $\text{Vol}(S)$ is the $d$-dimensional volume of set $S$, and $[0,a]$ denotes a $d$-dimensional box with $0$ and $a$ at opposite corners. $I_{x_i\in[0,a]}$ is the indicator function defined as $I=1$ for $x_i\in[0,a]$. Otherwise, $I=0$. The star discrepancy is
\EQQ
D_n^* = D_n^*(x_1, \cdots, x_n)=\sup_{a\in[0,1]^d} |\delta(a)|,
\ENN when $D_n^* \rightarrow 0$, then $\widehat{\mu}_n\rightarrow\mu$.
\edefinition}

For random numbers generated by QMC, their star discrepancy satisfies $D_n^*=O(n^{-1}\log(n)^{d-1})$ as $n\rightarrow \infty$. Thus, the convergence rate of QMC is $O(n^{-1+\varepsilon})$ for any $\varepsilon>0$, which is faster than that of MC which is $O(n^{-0.5})$ \cite{Owen2005}. Empirical comparisons demonstrate that QMC often outperforms MC for a reasonable sample number $n$.

{Due to the computational advantage of QMC, it is of interest to integrate it into MCMC in order to obtain a faster convergence rate. We call the resulted method as QMC-MCMC. It is easy to implement the QMC-MCMC by replacing the MC points by QMC points to generate the proposals and acceptances in an MCMC sampler. Details of the QMC-MCMC sampler is presented in Algorithm 1. We provide the theoretically consistency guarantee of QMC-MCMC as below. Before that, we need to introduce the completely uniformly distributed (CUD) sequence.}

\begin{algorithm}[t]
\caption{Proposed POPF Scheme}
\textbf{Step 1.} Given $D$ wind frams in a power system, collect wind speed data $\bm{x}=[x_1, x_2, \cdots, x_D]\t$ where $x_i$ with $i=1,2,\cdots, D$, represents the wind speed data of $i$-th wind farm.

\textbf{Step 2.} Obtain the joint PDF of the multivariate GMM
\EQQ
p(\bm{x}) = \sum_{m=1}^M c_m \mathcal{N}(\bm{x}|\bm{\mu}_m, \bm{\Sigma}_m),
\ENN with considering the correlation of $D$ wind farms. Where the parameter set $\Theta=\{c_m, \bm{\mu}_m, \bm{\Sigma}_m | m=1,2,\cdots,M\}$ is determined by EM algorithm.

\textbf{Step 3.} Draw wind speed samples from $p(\bm{x})$ via the QMC-MCMC sampler in Algorithm 1.

\textbf{Step 4.} Transform wind speed samples into wind power, then recursively calculate the DOPF problem (\ref{eq: POPF compact form}) with each wind power sample.

\textbf{Step 5.} Collect the output variable samples and computate their statistical information {such as} mean $\mu$ or STD $\sigma$.
\label{algo: POPF scheme}
\end{algorithm}

\bdefinition {\it (CUD)}
\label{def: CUD}
If for every integer $d\geq 1$, the points $z_i = (u_i, \cdots, u_{i+d-1})\in [0,1]^d$ satisfy $\lim_{n\rightarrow \infty} D_n^*(z_1, \cdots, z_n)=0$. Then the sequence $u_1, u_2, \cdots \in [0,1]$ is CUD.
\edefinition

{The Sobol sequence we employed in this paper is CUD. It can be integrated into the Metropolis Hastings algorithm to generate a consistent Markov chain as given in Lemma \ref{lemma: CUD on MCMC}. The detailed consistency proof is further explained in \cite{Owen2005} and \cite{Chentsov1967}.}

\blemma
\label{lemma: CUD on MCMC}
{Consider Markov chains with finite state spaces $\Omega = \{\omega_1, \cdots, \omega_K\}$, let $x_i\in \Omega$ for $i\geq 1$ be sampled from the standard construction for Markov chains, using a CUD sequence $u_i$. Assume that all $K^2$ transtion probabilities are positive.
Then $\widehat{\mu}_n\rightarrow \mu$ holds as $n\rightarrow\infty$ for all bounded $f$.}
\elemma

\begin{figure*}[htbp]
\centering
{\includegraphics[width=18cm]{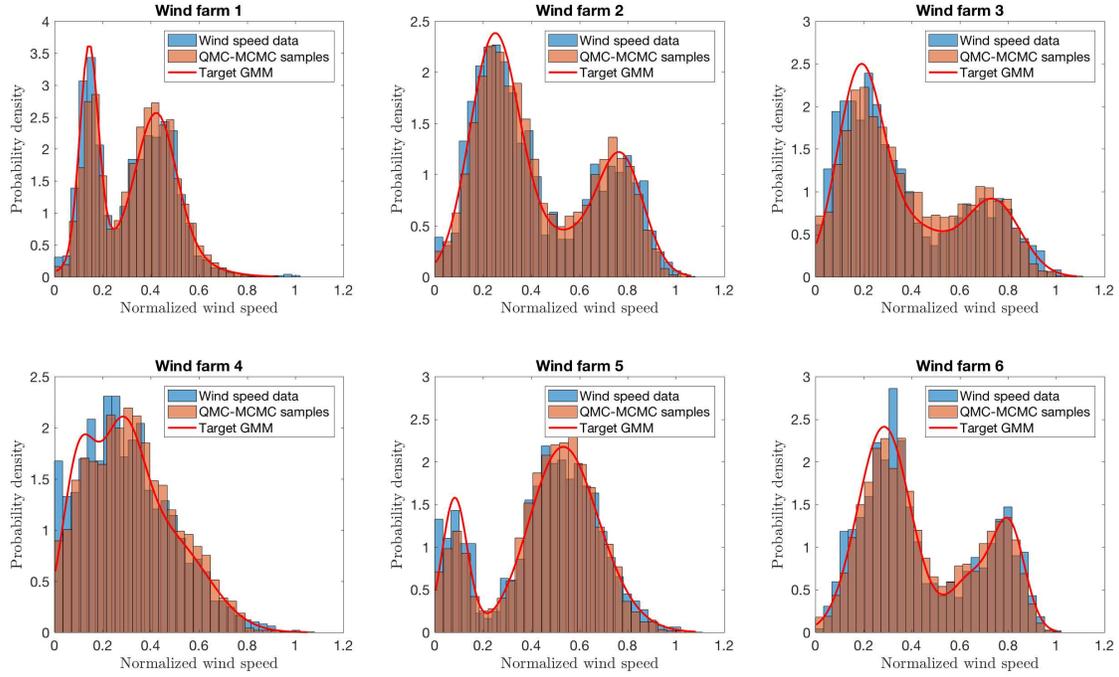}}
\caption{Target Gaussian mixture models of six wind farms and their samples by the proposed QMC-MCMC sampler.}
\label{fig: GMM and MCQMC samples}
\end{figure*}

\subsection{Proposed POPF Scheme}
\label{sec: prodedure}

{The proposed scheme for solving POPF with considering correlation of wind speeds with arbitrary distribution is summarized in Algorithm 2.
Note that in the last step of Algorithm 2, mean $\mu$ or standard deviation (STD) $\sigma$ is just rough analysis results with the output variable samples. If necessary, we can further establish the probability density function of output variable.}


\section{Case Studies}
\label{sec: case study}
{Case studies will be conducted on two benchmark systems, the modified IEEE 14-bus and 118-bus systems.} Eight wind turbines are integrated in one wind farm, the rated power of each wind turbine is {5 MW}. The rated, cut-in and cut-out wind speeds are set as 12 m/s, 2 m/s and 18 m/s \cite{Li2014}. The simulation was {conducted} on an MacBook Pro with 64-bit Intel i5 CPU at 2.3GHz and 8GB of RAM. MATPOWER, an MATLAB {power system simulation} {toolbox}, was adopted to solve the deterministic optimal power flow \cite{Zimmerman2011}.

The accuracy of POPF solving in this paper is estimated by calculating the errors of mean and stardand divation compared with the accuracy reference values. {POPF solutions using SRS with large enough sample size, e.g., $N=10000$, is set as the accurate reference.} The error index is defined as

\EQQ
\varepsilon_I^* = \Big|\frac{I_{a}^*-I_{s}^*}{I_{a}^*}\Big|\times 100\%.
\ENN Here $I$ is the statistical property such as the mean $\mu$ or STD $\sigma$ {associated with POPF {output variables}}. $I_{a}$ is the accurate reference value obtained from SRS, while $I_{s}$ is the simulated results using a certain sampling method with {$N$ samples, here $N\leq 10000$}. The symbol $*$ can be any output variables of the POPF computation such as the optimal cost, bus voltage or power flow.

\subsection{Modeling and Sampling Wind Speed}

The wind speed data of six wind farms is collected from the Measurement and Instrumentation Data Center under the National Renewable Energy Laboratory. {The wind speed data on mintuely basis for one day} at six different wind farms is used. Fig. \ref{fig: GMM and MCQMC samples} {illustrates histograms of the wind speed data in blue}. {It is observed that} the distributions of real wind speed data in one day can be arbitrary instead of following the Weibull distribution or any other known distributions.
The unspecific distributions of wind speed data motivate us to {deploy} GMM, to describe the probabilistic model of wind speed. As shown in Fig. \ref{fig: GMM and MCQMC samples}, the GMM obtained from six wind farms are {demonstrated} in red lines.
We do the unity-based normalization to cast the wind speed data into range $[0,1]$ for convenience of the algorithm {implementation}. Then, we {apply} the QMC-MCMC sampler in Algorithm 2 to generate samples from the six obtained Gaussian mixture models, as shown in Fig. \ref{fig: GMM and MCQMC samples} by orange. The distributions of samples in six wind farms are very similar to that of their original data.

\begin{table}[!]
\begin{center}
 \caption{Performance comparisons of SRS, LHS and QMC-MCMC on the modified IEEE 14-bus system with $N=2000$}
 \label{tab: error IEEE-14}
 \centering
 \begin{tabular}{|c|c|c|c|c|c|}
  \hline
  Variables & Methods & Mean & $\varepsilon_\mu(\%)$ & STD & $\varepsilon_\sigma(\%)$\\
  \hline
  \multirow{4}{*}{$Cost$($\$$)} & Ref & 12270 & $\backslash$ & 1225.1 & $\backslash$ \\
  \cline{2-6}
   & SRS-MCMC & 12295 & 0.2054 & 1286.1 & 4.9845 \\
  \cline{2-6}
  & LHS-MCMC & 12303 & 0.2706 & 1242.3 & 1.4075 \\
  \cline{2-6}
  & QMC-MCMC & 12270 & 0 & 1198.2 & 2.1940 \\
  \hline
  \multirow{4}{*}{$V$(p.u.)} & Ref & 1.0225 & $\backslash$ & 0.0220 & $\backslash$ \\
  \cline{2-6}
   & SRS-MCMC & 1.0232 & 0.0691 & 0.0213 & 3.0767 \\
  \cline{2-6}
  & LHS-MCMC & 1.0219 & 0.0654 & 0.0225 & 2.4378 \\
  \cline{2-6}
  & QMC-MCMC & 1.0222 & 0.0374 & 0.0226 & 2.5422 \\
  \hline
 \end{tabular}
\end{center}
\end{table}

\begin{figure}[htbp]
\centering
{\includegraphics[width=8cm]{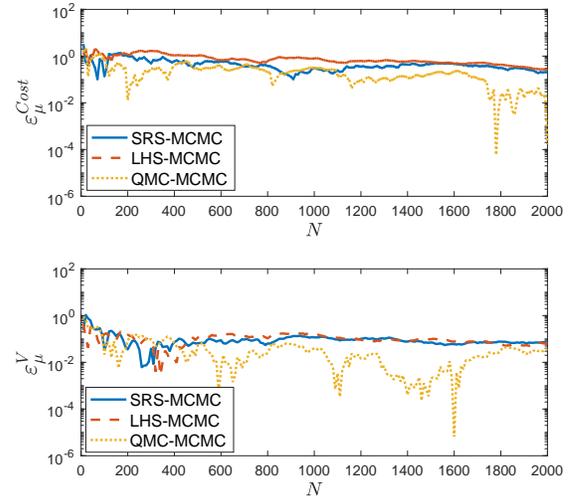}}
\caption{Performance comparisons of three sampling methods on the modified IEEE 14-bus system along increasing sampling size $N$.}
\label{fig: case14 error}
\end{figure}

\subsection{IEEE 14-bus System}

Three wind farms, the wind farm 1, 2, 3 are integrated at bus 3, 4, 5 of the IEEE 14-bus benchmark system, respectively. The conventional generator on bus 3 is removed and replaced by the wind farm 1. The rated capability of these three wind farms is {40 MW}. We assume the three wind farms in the modified IEEE 14-bus system are mutually influential, {the correlation among them is considered}.

We calculated the POPF results provided by {SRS with 10000 samples} as the accurate reference vaules. The MCMC sampling method is improved by the Latin hypercube sampling and Sobol-based QMC technique. Note that here we abbreviate the MCMC method with simple random sampling as SRS-MCMC, and MCMC improved by the Latin hypercube sampling and Sobol-based QMC technique as LHS-MCMC and QMC-MCMC, respectively. Performances of these three sampling methods with sampling size $N=2000$ are compared {against} the accurate reference vaules in Table~\ref{tab: error IEEE-14}. Note that the {voltage magnitude error index $\varepsilon_\mu^{V}$} in Table~\ref{tab: error IEEE-14} is measured at bus 12 in the IEEE 14-bus system. {It is observed that} the proposed QMC-MCMC method {almost} achieves relatively small mean error index $\varepsilon_\mu$ and STD error index $\varepsilon_\sigma$. To further compare their performance, we plot the mean error index of optimal cost $\varepsilon_\mu^{Cost}$ and voltage magnitude $\varepsilon_\mu^{V}$ {along increasing} sampling sizes in Fig. \ref{fig: case14 error}. The error index {associated with} QMC-MCMC method always keeps smaller than {those of} other two methods {which verifies the virtue of the former}.



\subsection{IEEE 118-bus System}
We now investigate the proposed POPF scheme on the IEEE 118-bus system \cite{Li2014}\cite{Xie2018}. As shown in Table \ref{tab: areas of IEEE-118}, the 118-bus system is divided into three areas {with each} area has been integrated into two wind farms. We consider the correlation between every two wind farms in each area. Specially, wind farms 1, 2 are integrated to bus 11, 17, wind farms 3, 4 are integrated to bus 37, 51, wind farms 5, 6 are integrated to bus 83, 96. The rated capability of these six wind farms is 40MW.

\begin{table}[!]
\begin{center}
 \caption{3 areas of IEEE 118-bus system with additional wind farms integrated.}
 \label{tab: areas of IEEE-118}
 \centering
 \begin{tabular}{|c|c|c|}
  \hline
  Areas & Buses & Buses with wind farms\\
  \hline
  1 & 1-23, 25-32, 113-115, 117 & 11, 17\\
  \hline
  2 & 33-69, 116 & 37, 51\\
  \hline
  3 & 24, 70-112, 118 & 83, 96\\
  \hline
 \end{tabular}
\end{center}
\end{table}

\begin{figure}[htbp]
\centering
{\includegraphics[width=8cm]{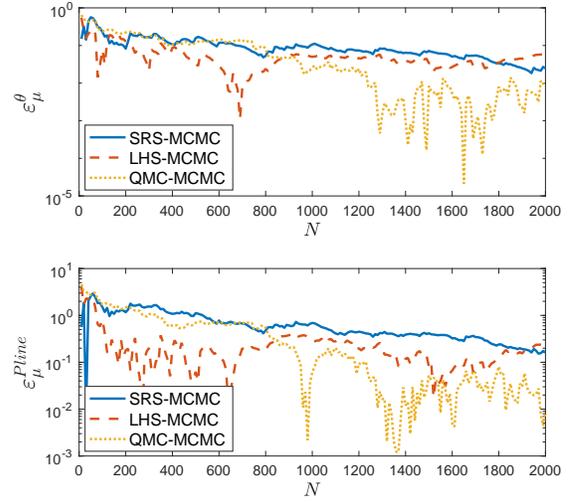}}
\caption{Performance comparisons of three sampling methods on the modified IEEE 118-bus system along increasing sampling size $N$.}
\label{fig: case118 error}\vspace{-2mm}
\end{figure}

Table \ref{tab: error IEEE-118} {provides} the performance comparisons of three sampling methods. Here, the Voltage angle $\theta$ at bus 98 and power flow $P_{line}$ at line 69-70 is measured. Mean error indeices of voltage angle $\varepsilon_\mu^{\theta}$ and power flow $\varepsilon_\mu^{P_{line}}$ with increasing sampling sizes are shown in Fig.\ref{fig: case118 error}. As can be seen, the QMC-MCMC method outperforms the other two ones.

\begin{table}[!]
\begin{center}
 \caption{Performance comparisons of SRS, LHS and QMC-MCMC on the modified IEEE 118-bus system with $N=2000$}
 \label{tab: error IEEE-118}
 \centering
 \begin{tabular}{|c|c|c|c|c|c|}
  \hline
  Variables & Methods & Mean & $\varepsilon_\mu(\%)$ & STD & $\varepsilon_\sigma(\%)$\\
  \hline
  \multirow{4}{*}{$\theta$(deg)} & Ref & 26.5042 & $\backslash$ & 0.5563 & $\backslash$ \\
  \cline{2-6}
   & SRS-MCMC & 26.5108 & 0.0248 & 0.5631 & 1.2265 \\
  \cline{2-6}
  & LHS-MCMC & 26.4891 & 0.0572 & 0.4789 & 13.9068 \\
  \cline{2-6}
    & QMC-MCMC & 26.5069 & 0.0100 & 0.5601 & 0.6959 \\
  \hline
  \multirow{4}{*}{$P_{line}$(MW)} & Ref & 2.4861 & $\backslash$ & 0.2708 & $\backslash$ \\
  \cline{2-6}
   & SRS-MCMC & 2.4821 & 0.1619 & 0.2740 & 1.1858 \\
  \cline{2-6}
  & LHS-MCMC & 2.4918 & 0.2287 & 0.2500 & 7.6757 \\
  \cline{2-6}
  & QMC-MCMC & 2.4859 & 0.0082 & 0.2707 & 0.0058 \\
  \hline
 \end{tabular}
\end{center}
\end{table}

\section{Conclusions}
\label{sec: Conclusions}
{In this paper, we developed a new stochastic scheme for POPF problem based on the multivariate GMM and Markov chain quasi-Monte Carlo sampling technique. The scheme approximates arbitrarily complex wind speed distribuions from multiple wind farms with considering their correlation. A MCMC sampler is adopted in our scheme to generate the wind speed samples for POPF solving. We novelly integrated the Sobol-based QMC technique into the MCMC sampling process to obtain a faster convergence rate. Two case studies on modified IEEE 14- and 118-bus systems with additional wind farms are conducted to verify the effectiveness of the proposed POPF scheme.}

\bibliographystyle{unsrt}
\small
\bibliography{mcmcpopf}

\begin{thebibliography}{10}

\bibitem{Wang2016}
S.~Wang, X.~Zhang, L.~Ge, and L.~Wu.
\newblock 2-{D} wind speed statistical model for reliability assessment of
  microgrid.
\newblock {\em IEEE Transactions on Sustainable Energy}, 7(3):1159--1169, 2016.

\bibitem{Chen2015}
Z.~Chen, L.~Wu, and M.~Shahidehpour.
\newblock Effective load carrying capability evaluation of renewable energy via
  stochastic long-term hourly based {SCUC}.
\newblock {\em IEEE Transactions on Sustainable Energy}, 6(1):188--197, 2015.

\bibitem{Wang2018}
S.~Wang, X.~Liu, K.~Wang, L.~Wu, and Y.~Zhang.
\newblock Tracing harmonic contributions of multiple distributed generations in
  distribution systems with uncertainty.
\newblock {\em International Journal of Electrical Power \& Energy Systems},
  95:585--591, 2018.

\bibitem{Li2014}
Y.~Li, W.~Li, W.~Yan, J.~Yu, and X.~Zhao.
\newblock Probabilistic optimal power flow considering correlations of wind
  speeds following different distributions.
\newblock {\em IEEE Transactions on Power Systems}, 29(4):1847--1854, 2014.

\bibitem{Ke2016}
D.~Ke, C.~Y. Chung, and Y.~Sun.
\newblock A novel probabilistic optimal power flow model with uncertain wind
  power generation described by customized {G}aussian mixture model.
\newblock {\em IEEE Transactions on Sustainable Energy}, 7(1):200--212, 2016.

\bibitem{Zhang2017}
Y.~Zhang, S.~Shen, and J.~L. Mathieu.
\newblock Distributionally robust chance-constrained optimal power flow with
  uncertain renewables and uncertain reserves provided by loads.
\newblock {\em IEEE Transactions on Power Systems}, 32(2):1378--1388, 2017.

\bibitem{Xie2018}
Z.~Q. Xie, T.~Y. Ji, M.~S. Li, and Q.~H. Wu.
\newblock Quasi-{Monte} {Carlo} based probabilistic optimal power flow
  considering the correlation of wind speeds using copula function.
\newblock {\em IEEE Transactions on Power Systems}, 33(2):2239--2247, 2018.

\bibitem{Kazemdehdashti2018}
A.~Kazemdehdashti, M.~Mohammadi, and A.~R. Seifi.
\newblock The generalized cross-entropy method in probabilistic optimal power
  flow.
\newblock {\em IEEE Transactions on Power Systems}, 2018.

\bibitem{Li2008}
X.~Li, Y.~Li, and S.~Zhang.
\newblock Analysis of probabilistic optimal power flow taking account of the
  variation of load power.
\newblock {\em IEEE Transactions on Power Systems}, 23(3):992--999, 2008.

\bibitem{Schellenberg2005}
A.~Schellenberg, W.~Rosehart, and J.~Aguado.
\newblock Cumulant-based probabilistic optimal power flow ({P-OPF}) with
  {G}aussian and gamma distributions.
\newblock {\em IEEE Transactions on Power Systems}, 20(2):773--781, 2005.

\bibitem{Zou2014}
B.~Zou and Q.~Xiao.
\newblock Solving probabilistic optimal power flow problem using quasi {Monte}
  {Carlo} method and ninth-order polynomial normal transformation.
\newblock {\em IEEE Transactions on Power Systems}, 29(1):300--306, 2014.

\bibitem{Aien2014}
M.~Aien, M.~Fotuhi-Firuzabad, and M.~Rashidinejad.
\newblock Probabilistic optimal power flow in correlated hybrid
  wind-photovoltaic power systems.
\newblock {\em IEEE Transactions on Smart Grid}, 5(1):130--138, 2014.

\bibitem{Verbic2006}
G.~Verbic and A.~Canizares.
\newblock Probabilistic optimal power flow in electricity markets based on a
  two-point estimate method.
\newblock {\em IEEE Transactions on Power Systems}, 21(4):1183--1893, 2006.

\bibitem{Schellenberg2005_1}
A.~Schellenberg, W.~Rosehart, and J.~A. Guado.
\newblock Introduction to cumulant-based probabilistic optimal power flow
  ({P-OPF}).
\newblock {\em IEEE Transactions on Power Systems}, 20(2):1184--1186, 2005.

\bibitem{Wei2018}
W.~Wei, J.~Wang, and L.~Wu.
\newblock Distribution optimal power flow with real-time price elasticity.
\newblock {\em IEEE Transactions on Power Systems}, 33(1):1097--1098, 2018.

\bibitem{Su2005}
C.~L. Su.
\newblock Probabilistic load-flow computation using point estimate method.
\newblock {\em IEEE Transactions on Power Systems}, 20(4):1843--1851, 2005.

\bibitem{Yu2009}
H.~Yu, C.~Y. Chung, K.~P. Wong, H.~W. Lee, and J.~H. Zhang.
\newblock Probabilistic load flow evaluation with hybrid {L}atin hypercube
  sampling and {C}holesky decomposition.
\newblock {\em IEEE Transactions on Power Systems}, 24(2):661--667, 2009.

\bibitem{Hajian2013}
M.~Hajian, W.~D. Rosehart, and H.~Zareipour.
\newblock Probabilistic power flow by {M}onte {C}arlo simulation with {L}atin
  supercube sampling.
\newblock {\em IEEE Transactions on Power Systems}, 28(2):1550--1559, 2013.

\bibitem{Xu2017}
X.~Y. Xu and Z.~Yan.
\newblock Probabilistic load flow calculation with quasi-{M}onte {C}arlo and
  multiple linear regression.
\newblock {\em International Journal of Electrical Power \& Energy System},
  88:1--12, 2017.

\bibitem{Wang2017a}
Z.~Wang, C.~Shen, F.~Liu, and F.~Gao.
\newblock Analytical expressions for joint distributions in probabilistic load
  flow.
\newblock {\em IEEE Transactions on Power Systems}, 32(3):2473--2474, 2017.

\bibitem{Fan2012}
M.~Fan, V.~Vittal, G.~T. Heydt, and R.~Ayyanar.
\newblock Probabilistic power flow studies for transmission systems with
  photovoltaic generation using cumulants.
\newblock {\em IEEE Transactions on Power Systems}, 27(4):2251--2261, 2012.

\bibitem{Williams2013}
T.~Williams and C.~Crawford.
\newblock Probabilistic load flow modeling comparing maximum entropy and
  {G}ram-{C}harlier probability density function reconstructions.
\newblock {\em IEEE Transactions on Power Systems}, 28(1):272--280, 2013.

\bibitem{Singh2010}
R.~Singh, B.~C. Pal, and R.~A. Jabr.
\newblock Statistical representation of distribution system loads using
  {G}aussian mixture model.
\newblock {\em IEEE Transactions on Power Systems}, 25(1):29--37, 2010.

\bibitem{Wang2017}
Z.~Wang, C.~Shen, F.~Liu, X.~Wu, C.~C. Liu, and F.~Gao.
\newblock Chance-constrained economic dispatch with non-{G}aussian correlated
  wind power uncertainty.
\newblock {\em IEEE Transactions on Power Systems}, 32(6):4880--4893, 2017.

\bibitem{Tang2017}
Y.~Tang, K.~Dvijotham, and S.~Low.
\newblock Real-time optimal power flow.
\newblock {\em IEEE Transactions on Smart Grid}, 8(6):2963--2973, 2017.

\bibitem{Patel2005}
M.~R. Patel.
\newblock {\em Wind and solar power systems: design, analysis, and operation}.
\newblock CRC press, 2005.

\bibitem{Gan2014}
L.~Gan and H.~L. Steven.
\newblock Optimal power flow in direct current networks.
\newblock {\em IEEE Transactions on Power Systems}, 29(6):2892--2904, 2014.

\bibitem{Ross2014}
S.~M. Ross.
\newblock {\em {I}ntroduction to {P}robability {M}odels}.
\newblock Academic press, 2014.

\bibitem{Robert2014}
C.~Robert.
\newblock {\em {M}achine {L}earning: {A} {P}robabilistic {P}erspective}.
\newblock 2014.

\bibitem{Papaefthymiou2008}
G.~Papaefthymiou and B.~Klockl.
\newblock {MCMC} for wind power simulation.
\newblock {\em IEEE Transactions on Energy Conversion}, 23(1):234--240, 2008.

\bibitem{Gilks1995}
R.~Sylvia W.~R.~Gilks and S.~David.
\newblock {\em {M}arkov {C}hain {M}onte {C}arlo in {P}ractice}.
\newblock CRC press, 1995.

\bibitem{Dick2013}
J.~Dick, F.~Y. Kuo, and I.~H. Sloan.
\newblock High-dimensional integration: the quasi-{M}onte {C}arlo way.
\newblock {\em Acta Numerica}, 22:133--288, 2013.

\bibitem{Bratley1988}
P.~Bratley and B.~L. Fox.
\newblock Algorithm 659: Implementing {S}obol's quasirandom sequence generator.
\newblock {\em ACM Transactions on Mathematical Software (TOMS)},
  14(1):88--100, 1988.

\bibitem{Niederreiter1992}
H.~Niederreiter.
\newblock {\em Random number generation and quasi-{M}onte {C}arlo methods}.
\newblock {SIAM}, 1992.

\bibitem{Owen2005}
A.~B. Owen and S.~D. Tribble.
\newblock A quasi-{M}onte {C}arlo {M}etropolis algorithm.
\newblock {\em Proceedings of the National Academy of Sciences},
  102(25):8844--8849, 2005.

\bibitem{Chentsov1967}
N.~N. Chentsov.
\newblock Pseudorandom numbers for modelling {M}arkov chains.
\newblock {\em USSR Computational Mathematics and Mathematical Physics},
  7(3):218--233, 1967.

\bibitem{Zimmerman2011}
R.~D. Zimmerman, C.~E. Murillo-Sanchez, and J.~T. Robert.
\newblock {MATPOWER}: Steady-state operations, planning, and analysis tools for
  power systems research and education.
\newblock {\em IEEE Transactions on Power Systems}, 26(1):12--19, 2011.

\end{thebibliography}

\end{document}